\documentclass[pra,twocolumn,showpacs,preprintnumbers,superscriptaddress]{revtex4}
\usepackage{times}
\usepackage{bm}
\usepackage{graphicx}
\usepackage{amsbsy}
\usepackage{amsmath}
\usepackage{amsfonts}
\usepackage{amsthm}
\usepackage{xcolor}

\begin{document}

\theoremstyle{plain}
\newtheorem{theorem}{Theorem}
\newtheorem{lemma}[theorem]{Lemma}
\newtheorem{corollary}[theorem]{Corollary}
\newtheorem{proposition}[theorem]{Proposition}\newtheorem{conjecture}[theorem]{Conjecture}
\theoremstyle{definition}
\newtheorem{definition}[theorem]{Definition}
\theoremstyle{remark}
\newtheorem*{remark}{Remark}
\newtheorem{example}{Example}
\title{Construction of a Family of Positive But Not Completely Positive Map For the Detection of Bound Entangled States}
\author{Richa Rohira, Shreya Sanduja, Satyabrata Adhikari}
\email{richarohira_2k19mscmat12@dtu.ac.in, shreyasanduja_2k19mscmat10@dtu.ac.in, satyabrata@dtu.ac.in} \affiliation{Delhi Technological University, Shahbad Daulatpur, Main Bawana Road, Delhi-110042,
India}

\begin{abstract}
We construct a family of map which is shown to be positive when imposing certain condition on the parameters. Then we show that the constructed map can never be completely positive. After tuning the parameters, we found that the map still remain positive but it is not completely positive. We then use the positive but not completely positive map in the detection of bound entangled state and negative partial transpose entangled states.
\end{abstract}
 \maketitle
%Mathematics Subject Classification: 81P45, 15A60 \maketitle
\section{Introduction}
Entanglement, first introduced in the EPR paper \cite{einstein}, is a quantum mechanical feature that can be used as a resource for computational and communicational purposes \cite{ekert1,horodecki4}. It plays a central role in many information processing protocols such as quantum cryptography \cite{ekert}, quantum superdense coding \cite{bennett2} and quantum teleportation \cite{bennett1,bennett3}. The potential offered by quantum entanglement to computing, security and communication makes it a topic of vital interest to researchers all across the globe.\\
One of the important problem in quantum information theory is the detection of entanglement in a quantum mechanical system. A pure two-qubit entangled system always violated Bell-CHSH inequality and thus detected by Bell-CHSH operator \cite{bennett4,deutsch}. On the other hand, the Bell-CHSH inequality fails to detect the several mixed bipartite entangled state. This loophole can be fixed using Peres-Horodecki (PH) positive partial transpose (PPT) criteria, which is necessary and sufficient for the detection of entanglement in $2 \otimes 2$ and $2 \otimes 3$ systems \cite{peres,horo1}. In higher dimensional systems, all states with negative partial transpose (NPT) are entangled but the states with positive partial transpose may or may not be entangled \cite{phorodecki}. The entangled states which are described by a density matrix that remains positive under partial transposition are known as bound entangled states. Thus, the separability problem can also be framed as analysing whether states with positive partial transposition are entangled or not.\\
The separability problem can be tackled to certain extent by witness operator \cite{lewenstein,ganguly}. Witness operators are hermitian operators with at least one negative eigenvalues. The witness operators are more powerful than Bell inequalities in the sense that it can detect multipartite entanglement in different cuts, if some prior information about the state under investigation is provided. They not only detect multipartite entanglement in different cuts but also detect genuine entanglement and classify entanglement in multipartite system. They are observables and thus experimentally realizable also.\\
A map $\Lambda:M_{d_{1}}(C)\rightarrow M_{d_{2}}(C)$ is said to be positive map if $\Lambda(A) \in M_{d_{2}}(C)$ is positive, for any positive $A \in M_{d_{1}}(C)$. Unfortunately, the structure of positive map is not completely understood and still it is under extensive research \cite{stormer1,stormer2,woronowicz}. Indecomposable maps plays an important role among all positive linear maps due to the fact that it can detect positive partial transpose entangled states. Let $\Lambda$ be a positive map and let $I_{d}:M_{d}(C)\rightarrow M_{d}(C)$ denote an identity map. Then we say that $\Lambda$ is completely positive if for all d, the extended map $I_{d} \otimes \Lambda$ is positive. Trace map is an example of a completely positive map. There also exist a map such as transposition map which is positive but not completely positive map. These positive but not completely positive maps are important in detecting the entanglement in a composite quantum system.\\
The separability problem can be reformulated in terms of positive maps \cite{horo1} as follows: Let us suppose that $H_{d_{1}}$ and $H_{d_{2}}$ represent two Hilbert spaces with dimensions $d_{1}$ and $d_{2}$ respectively. A bipartite quantum state described by the density operator $\sigma \in H_{d_{1}}\otimes H_{d_{2}}$ is separable if and only if $(I_{d_{1}} \otimes \Lambda)\sigma$ is positive for any positive map $\Lambda$. Thus there is a deep relation between the theory of detection of entanglement and operator theory. This linkage has been established by Choi-Jamiolkowski isomorphism \cite{choi,jamiolkowski}. According to Choi-Jamiolkowski isomorphism, there is a one to one correspondence  between entanglement witnesses and a positive but not completely positive map.\\
The motivation of this work is as follows: The construction and studying the structure of new positive map may give useful insight in the understanding of positive map, which gives us the first motivation of this work. Secondly, we find that the problem of constructing the positive but not completely positive map and its relation in the detection of entanglement may take one step further in the development of not only operator theory but also quantum information theory.\\
The work is organized as follows: In section-II, we review few earlier results that is needed in our work. In section-III, we have constructed a family of map and then characterize it for when the map is (i) positive or (ii) completely positive or (iii) positive but not completely positive. In section-IV, we have chosen a particular map from the class of positive but not completely positive map and then shown that it can detect bound entangled entangled state. In section-V, we end up with the conclusion.
\section{Preliminaries}
\textbf{Definition-1:} A matrix $V$ is said to be a contraction if
\begin{equation}
\|V\|\leq 1
\label{contraction}
\end{equation}
where $\|.\|$ denote the operator norm.\\
\textbf{Result-1 \cite{stormer}:} Let $\phi: M_{m}(C)\rightarrow M_{n}(C)$ be a linear map. Then the following statements are equivalent:\\
(i) $\phi$ is completely positive.\\
(ii) $C_{\phi}$ is positive semidefinite, where $C_{\phi}$ denotes the Choi matrix of $\phi$.\\

\textbf{Result-2 \cite{zhang}:} Let A be a partitioned block matrix of the form
\begin{eqnarray} A=
 \begin{pmatrix}
 X & Y \\
 Y^{*} & Z
 \end{pmatrix}
 \label{block}
\end{eqnarray}
Then A is positive semidefinite if and only if X and Z are positive semidefinite and there exists a contraction $V$ such that
\begin{equation}
Y = X^{\frac{1}{2}}VZ^{\frac{1}{2}}
\label{psdcond1}
\end{equation}
\textbf{Result-3 \cite{sharma}:} For positive definite blocks $X$ and $Z$, the matrix $A$ given in (\ref{block}) is positive semidefinite iff
$X\geq YZ^{-1}Y^{\dagger}$.
\section{Construction of a family of map}
In this section, we will construct a map and derive the condition for which the map is positive. Further, we will probe that whether the constructed map is completely positive. Moreover, we will provide the explicit matrix form of the map, which is positive but not completely positive.\\
Let us take a positive integer $n~~(n\geq 2)$ and then we define a general family of map $\Phi: M_{n}(C)\rightarrow M_{n}(C) \otimes M_{n}(C)$ as
\begin{equation}
\Phi_{\alpha,\beta}(A) = \alpha((A + A^{T}) \otimes I_{n}) + \beta ( |\psi_{+}\rangle\langle \psi_{+}| )^{\Gamma}
\label{map}
\end{equation}
where $A$ denote $n\times n$ matrix, $\alpha,\beta \in R$, $\Gamma$ represent the partial transposition, $I_{n}$ denote the identity matrix of order $n$ and $|\psi_{+}\rangle=\frac{1}{\sqrt{n}}\sum_{i=1}^{n}|ii\rangle$.\\
To discuss our result, we will fix $n=2$ and re-define the map $\Phi: M_{2}(C)\rightarrow M_{2}(C) \otimes M_{2}(C)$ as
\begin{equation}
\Phi_{\alpha,\beta}(A) = \alpha((A + A^{T}) \otimes I_{2}) + \beta ( |\psi_{+}\rangle\langle \psi_{+}| )^{\Gamma}
\label{map}
\end{equation}
where $I_{2}$ denote the identity matrix of order $2$ and $|\psi_{+}\rangle=\frac{1}{\sqrt{2}}(|00\rangle + |11\rangle)$.\\
For any $a,d \geq 0$ and $b,c \in \textrm{R}$, we can take the input matrix $A \in M_{2}(R)$ of the form
\begin{eqnarray} A=
 \begin{pmatrix}
 a & b \\
 c & d
 \end{pmatrix}
 \label{input}
\end{eqnarray}
In matrix notation, the output of the map $\Phi_{\alpha,\beta}$ can be expressed as
\begin{eqnarray}
\Phi_{\alpha,\beta}=
 \begin{pmatrix}
  2a\alpha + \frac{\beta}{2}& 0 & \alpha(b+c) & 0  \\
     0  &  2a\alpha & \frac{\beta}{2} & \alpha(b+c)   \\
      \alpha(b+c) & \frac{\beta}{2} & 2d\alpha  & 0 \\
    0 & \alpha(b+c) & 0 & 2d\alpha + \frac{\beta}{2}
  \end{pmatrix}
  \label{mapmat}
\end{eqnarray}
\subsection{Conditions for which a map $\phi$ will be positive}
We will derive here the conditions for which $\Phi$ represent a positive map. The map $\Phi$ will be positive if the matrix represented by $\Phi_{\alpha,\beta}$ given in (\ref{mapmat}) is a positive semi-definite matrix. To accomplish this task, we re-express $\Phi_{\alpha,\beta}$ in a block matrix form as
\begin{eqnarray}
\Phi_{\alpha,\beta}=
\begin{pmatrix}
 X & Y \\
 Y^{*} & Z
 \end{pmatrix}
\label{block1}
\end{eqnarray}
where
\begin{eqnarray}
&&X=
\begin{pmatrix}
 2a\alpha + \frac{\beta}{2}& 0 \\
 0  &  2a\alpha
 \end{pmatrix}, Y=\begin{pmatrix}
 \alpha(b+c) & 0  \\
 \frac{\beta}{2} & \alpha(b+c)
 \end{pmatrix},\nonumber\\&&Z=\begin{pmatrix}
 2d\alpha  & 0 \\
 0 & 2d\alpha + \frac{\beta}{2}
 \end{pmatrix}
\label{block1}
\end{eqnarray}
Applying Result-2 on $\Phi_{\alpha,\beta}$, we can state that the matrix $\Phi_{\alpha,\beta}$ will be positive semidefinite if the following conditions hold:\\
\begin{eqnarray}
(i) X\geq 0\Rightarrow 2a\alpha \geq 0 ~~\textrm{and}~~ 4a\alpha + \beta \geq 0
\label{cond1}
\end{eqnarray}
\begin{eqnarray}
(ii) Z\geq 0\Rightarrow 2d\alpha \geq 0 ~~\textrm{and}~~ 4d\alpha + \beta \geq 0
\label{cond2}
\end{eqnarray}
\begin{eqnarray}
(iii)&& \|V\|=\|X^{\frac{-1}{2}}YZ^{\frac{-1}{2}}\|\leq 1 \nonumber\\&& \Rightarrow
\|\begin{pmatrix}
\frac{\alpha(b+c)}{\sqrt{d\alpha(4a\alpha+\beta)}} & 0\\
 \frac{\beta}{4\alpha\sqrt{ad}}  & \frac{\alpha(b+c)}{\sqrt{a\alpha(4d\alpha+\beta)}}
 \end{pmatrix}\|\leq 1
\label{cond3}
\end{eqnarray}
where $\|V\|$ denote the operator norm of $V$.\\
Conditions $(i)$ and $(ii)$ given by (\ref{cond1}) and (\ref{cond2}) are collectively given by
\begin{eqnarray}
2\alpha(a+d) +\beta \geq 0,~~\alpha \geq 0
\label{cond100}
\end{eqnarray}
Now our task is to take into account condition (iii) in which we need to calculate the operator norm of the matrix $V$. Operator norm of the matrix $V$ is defined as the maximum eigenvalue of $V^{T}V$. The eigenvalue of $V^{T}V$ can be calculated from the characteristic equation of $V^{T}V$. The characteristic equation of $V^{T}V$ is given by
\begin{eqnarray}
&&\lambda^2 - k_{1}\lambda + \frac{k_{2}}{4}=0
\label{char}
\end{eqnarray}
where $k_{1}=\frac{\alpha(b+c)^2}{(4a\alpha+\beta)d} + \frac{\beta}{4a\alpha} + \frac{\beta^2}{16ad\alpha^2} + \frac{d}{a}$ and $k_{2}=\frac{(b+c)^2 (\beta+4d\alpha)}{ad(4a\alpha+\beta)}$.\\
Since $a\geq 0$ and $d\geq0$ from the earlier assumptions and using equations (\ref{cond1}) and (\ref{cond2}), we can infer that $k_{1}\geq 0$ and $k_{2}\geq 0$. Thus, it is clear from Descarte's rule of sign that the two roots of the characteristic equation given by (\ref{char}) will be positive.
If $\lambda_{1}$ and $\lambda_{2}$ denote two positive eigenvalues of $V^{T}V$ then they are given by
\begin{eqnarray}
&&\lambda_1 = \frac{1}{2}(k_{1} + \sqrt{k_{1}^{2} - k_{2}})\nonumber\\&&
\lambda_2 = \frac{1}{2}(k_{1} - \sqrt{k_{1}^{2} - k_{2}})
\label{eigen1}
\end{eqnarray}
Since both the eigenvalues are positive so $\|V\|=max\{\lambda_{1},\lambda_{2}\}=\lambda_{1}$. The condition (iii) says that $\|V\|\leq 1$ which implies
\begin{eqnarray}
4(1 + \sqrt{k_{1}^{2} - k_{2}} -k_{2}) \geq 1
\label{cond300}
\end{eqnarray}
The map $\Phi_{\alpha,\beta}$ is positive if equations (\ref{cond100}) and (\ref{cond300}) holds simultaneously.
In particular, the map $\Phi_{\alpha,\beta}$ will be positive for $\alpha\geq 0$ and $\beta=0$.
\subsection{Is the map $\phi$ completely positive?}
In this section, we will investigate the fact that whether the map $\phi$ is completely positive. To do this, we begin with the construction of Choi matrix corresponding to the positive operator $\Phi_{\alpha,\beta}$. The Choi matrix $C_{\Phi_{\alpha,\beta}}$ is defined as \cite{choi}
\begin{eqnarray}
C_{\Phi_{\alpha,\beta}}=\sum_{i,j=0}^{1}|i\rangle\langle j|\otimes \Phi_{\alpha,\beta}(|i\rangle\langle j|)
\label{choimatrix}
\end{eqnarray}
where $|i\rangle$ represent the basis state in two-dimensional Hilbert space.\\
The Choi matrix $C_{\Phi_{\alpha,\beta}}$ can be re-expressed in terms of matrix as
\begin{eqnarray}
C_{\Phi_{\alpha,\beta}}=
 \begin{pmatrix}
  2\alpha  +\frac{\beta}{2} & 0 & 0 & 0 & \frac{\beta}{2} & 0 & \alpha & 0 \\
  0 & 2\alpha & \frac{\beta}{2} & 0 & 0 & 0 & \frac{\beta}{2} & \alpha \\
  0 & \frac{\beta}{2} & 0 & 0 & \alpha & \frac{\beta}{2} & 0 & 0\\
0 & 0 & 0 & \frac{\beta}{2} & 0 & \alpha & 0 & \frac{\beta}{2} \\
\frac{\beta}{2} & 0 & \alpha & 0 & \frac{\beta}{2} & 0 & 0 & 0\\
0 & 0 & \frac{\beta}{2} & \alpha & 0 & 0 & \frac{\beta}{2} & 0\\
\alpha & \frac{\beta}{2} & 0 & 0 & 0 & \frac{\beta}{2} & 2\alpha  & 0\\
0 & \alpha & 0 & \frac{\beta}{2} & 0 & 0 & 0 & 2\alpha  +\frac{\beta}{2}
  \end{pmatrix},
  \label{choimatrix1}
\end{eqnarray}
To show the completely positivity of a positive map $\Phi_{\alpha,\beta}$, we need to the show that the choi matrix $C_{\Phi_{\alpha,\beta}}$ corresponding to the positive map $\Phi_{\alpha,\beta}$ is positive semidefinite. We first express the choi matrix in block form as
\begin{eqnarray}
C_{\Phi_{\alpha,\beta}}=
\begin{pmatrix}
 P & Q \\
 Q^{*} & R
 \end{pmatrix}
\label{block1}
\end{eqnarray}
where
\begin{eqnarray}
&&P=
\begin{pmatrix}
 2\alpha  +\frac{\beta}{2} & 0 & 0 & 0 \\
 0 & 2\alpha  & \frac{\beta}{2} & 0 \\
 0 & \frac{\beta}{2} & 0 & 0 \\
 0 & 0 & 0 & \frac{\beta}{2}
 \end{pmatrix}, Q=\begin{pmatrix}
 \frac{\beta}{2} & 0 & \alpha & 0 \\
 0 & 0 & \frac{\beta}{2} & \alpha \\
 \alpha & \frac{\beta}{2} & 0 & 0 \\
 0 & \alpha & 0 & \frac{\beta}{2}
 \end{pmatrix},\nonumber\\&& R=\begin{pmatrix}
 \frac{\beta}{2} & 0 & 0 & 0 \\
 0 & 0 & \frac{\beta}{2} & 0 \\
 0 & \frac{\beta}{2} & 2\alpha & 0 \\
 0 & 0 & 0 & 2\alpha  + \frac{\beta}{2}
 \end{pmatrix}
\label{block2}
\end{eqnarray}
Following Result-3, we can show that the choi matrix $C_{\Phi_{\alpha,\beta}}$ is positive semidefinite if and only if the following conditions are satisfied:
\begin{eqnarray}
(i)~~ P\geq 0 ~~\textrm{holds when}~~\beta  = 0~~\textrm{and}~~ \alpha  \ge 0
\label{cond12}
\end{eqnarray}
\begin{eqnarray}
(ii)~~ R \geq 0~~\textrm{holds when}~~\beta  = 0~~ \textrm{and}~~ \alpha  \ge 0
\label{cond22}
\end{eqnarray}
\begin{eqnarray}
&&(iii)~~P - Q{R^{ - 1}}Q^{*} \ge 0~~ \textrm{holds for}~~ \textrm{either}~~\nonumber\\&& (\alpha=0~~ \textrm{and}~~ \beta \neq 0)~~ \textrm{or}~~ (\alpha  > 0~~ \textrm{and}~~ 4\alpha +\beta  < 0)\nonumber\\&&~~ \textrm{or} ~~ (\alpha  > 0~~,~~ 3\alpha +2\beta  \ge 0~~\textrm{and}~~\beta \neq 0)
\label{cond23}
\end{eqnarray}
It can be easily observe that the conditions $(i)$, $(ii)$ and $(iii)$ does not hold simultaneously. Thus the map $\Phi_{\alpha,\beta}$ is not completely positive.
\subsection{Conditions for which a map $\phi$ will be positive but not completely positive}
In the previous sections, we have derived the condition for which the map $\Phi$ will be positive and later we proved that the positive map $\Phi$ cannot be completely positive. In this section, we will derive the common interval of $\alpha$ for which the map $\Phi$ will be positive but not completely positive simultaneously.\\
Without any loss of generality, let us consider the $2\times 2$ positive matrix $A_{1}\in M_{2}(\textrm{R})$ as
\begin{eqnarray}
A_{1}= \begin{pmatrix}
 \frac{1}{4} & \frac{1}{3} \\
 \frac{1}{9} & 2
\end{pmatrix}
\end{eqnarray}
Further, taking $\beta=-\gamma (\gamma>0)$, the output of the mapping can be represented by the matrix as
\begin{eqnarray}
\Phi_{\alpha,-\gamma}(A_{1})=
  \begin{pmatrix}
  \frac{\alpha}{2} - \frac{\gamma}{2}& 0 & \frac{4\alpha}{9} & 0  \\
     0  &  \frac{\alpha}{2} & -\frac{\gamma}{2} & \frac{4\alpha}{9}   \\
      \frac{4\alpha}{9} & -\frac{\gamma}{2} & 4\alpha  & 0 \\
    0 & \frac{4\alpha}{9} & 0 & 4\alpha - \frac{\gamma}{2}
  \end{pmatrix}
  \label{mapmat300}
\end{eqnarray}
It can be easily shown that the map $\Phi_{\alpha,-\gamma}$ always produces a positive matrix at the output if $\gamma>0$ and $\alpha\geq \frac{9\gamma}{2\sqrt{146}}$. Thus $\Phi_{\alpha,-\gamma}$ represent a positive map if $\gamma>0$ and $\alpha\geq \frac{9\gamma}{2\sqrt{146}}$.  Furthermore, the Choi matrix corresponding to the positive map $\Phi_{\alpha,-\gamma}$ is given by
\begin{eqnarray}
C_{\Phi_{\alpha,-\gamma}}=
 \begin{pmatrix}
  m_{1} & 0 & 0 & 0 & -\frac{\gamma}{2} & 0 & \alpha & 0 \\
  0 & 2\alpha & -\frac{\gamma}{2} & 0 & 0 & 0 & -\frac{\gamma}{2} & \alpha \\
  0 & -\frac{\gamma}{2} & 0 & 0 & \alpha & -\frac{\gamma}{2} & 0 & 0\\
0 & 0 & 0 & -\frac{\gamma}{2} & 0 & \alpha & 0 & -\frac{\gamma}{2} \\
-\frac{\gamma}{2} & 0 & \alpha & 0 & -\frac{\gamma}{2} & 0 & 0 & 0\\
0 & 0 & -\frac{\gamma}{2} & \alpha & 0 & 0 & -\frac{\gamma}{2} & 0\\
\alpha & -\frac{\gamma}{2} & 0 & 0 & 0 & -\frac{\gamma}{2} & 2\alpha & 0\\
0 & \alpha & 0 & -\frac{\gamma}{2} & 0 & 0 & 0 & m_{1}
  \end{pmatrix}
  \label{choimatrix100}
\end{eqnarray}
where $m_{1}=2\alpha-\frac{\gamma}{2}$.\\
The eigenvalues of $C_{\Phi_{\alpha,-\gamma}}$ are given by
\begin{eqnarray}
&&\mu_{1}=\frac{-\gamma+\sqrt{4\alpha^{2}+\gamma^{2}}}{2},~~\mu_{2}=\frac{-\gamma-\sqrt{4\alpha^{2}+\gamma^{2}}}{2}\nonumber\\&&
\mu_{3}=\frac{4\alpha-\gamma+\sqrt{4\alpha^{2}+\gamma^{2}}}{2},\nonumber\\&&
\mu_{4}=\frac{4\alpha-\gamma-\sqrt{4\alpha^{2}+\gamma^{2}}}{2},\nonumber\\&&
\mu_{5}=\alpha+\sqrt{\frac{4\alpha^{2}+\gamma^{2}+\sqrt{16\alpha^{2}+4\alpha^{2}\gamma^{2}+\gamma^{4}}}{2}}\nonumber\\&&
\mu_{6}=\alpha+\sqrt{\frac{4\alpha^{2}+\gamma^{2}-\sqrt{16\alpha^{2}+4\alpha^{2}\gamma^{2}+\gamma^{4}}}{2}}\nonumber\\&&
\mu_{7}=\alpha-\sqrt{\frac{4\alpha^{2}+\gamma^{2}+\sqrt{16\alpha^{2}+4\alpha^{2}\gamma^{2}+\gamma^{4}}}{2}}\nonumber\\&&
\mu_{8}=\alpha-\sqrt{\frac{4\alpha^{2}+\gamma^{2}-\sqrt{16\alpha^{2}+4\alpha^{2}\gamma^{2}+\gamma^{4}}}{2}}\nonumber\\&&
\label{cond320}
\end{eqnarray}
 It can be observed that the Choi matrix $C_{\Phi_{\alpha,-\gamma}}$ has at least one negative eigenvalues for any $\alpha$ and $\gamma$. Therefore, $\Phi_{\alpha,-\gamma}$ is not completely positive map for any $\alpha$ and $\gamma$. Thus for $\gamma>0$ and $\alpha\geq \frac{9\gamma}{2\sqrt{146}}$, the map $\Phi_{\alpha,-\gamma}$ is positive but not completely positive.
\section{Positive but not completely positive map act as witness operator for the detection of entangled states}
In this section, we will construct a specific map which is positive but not completely positive and then use it to detect negative partial transpose entangled states and bound entangled states. We will construct the Choi matrix from the positive map that can be considered as a witness operator. A witness operator $W$ is a hermitian operator, which satisfies the following properties:
\begin{eqnarray}
&&(i)~~Tr(W \rho_{s})\geq 0,~~ \textrm{for all separable state}~~ \rho_{s}\nonumber\\&&
(ii)~Tr(W \rho_{e})<0,~~ \textrm{for at least one entangled state}~~ \rho_{e}\nonumber\\&&
\label{witnesscond}
\end{eqnarray}
\subsection{Detection of Bound Entangled State}
To achieve our task, let us first fix $\gamma=2$ and then choose a value of $\alpha$ from the interval $\alpha\geq \frac{9}{\sqrt{146}}$. Taking $\alpha=\frac{3}{4}$, the matrix given in (\ref{mapmat300}) reduces to
\begin{eqnarray}
\Phi_{\frac{3}{4},-2}(A_{1})=
  \begin{pmatrix}
  \frac{11}{8}& 0 & \frac{1}{3} & 0  \\
     0  &  \frac{3}{8} & -1 & \frac{1}{3}   \\
      \frac{1}{3} & -1 & 3 & 0 \\
    0 & \frac{1}{3} & 0 & 4
  \end{pmatrix}
  \label{mapmat301}
\end{eqnarray}
In particular, the map $\Phi_{\frac{3}{4},-2}$ represent a positive map.  Using this positive map, we can construct the Choi matrix which is given below:
\begin{eqnarray}
C_{\Phi_{\frac{3}{4},-2}}=
 \begin{pmatrix}
  \frac{1}{2} & 0 & 0 & 0 & -1 & 0 & \frac{3}{4} & 0 \\
  0 & \frac{3}{2} & -1 & 0 & 0 & 0 & -1 & \frac{3}{4} \\
  0 & -1 & 0 & 0 & \frac{3}{4} & -1 & 0 & 0\\
0 & 0 & 0 & -1 & 0 & \frac{3}{4} & 0 & -1 \\
-1 & 0 & \frac{3}{4} & 0 & -1 & 0 & 0 & 0\\
0 & 0 & -1 & \frac{3}{4} & 0 & 0 & -1 & 0\\
\frac{3}{4} & -1 & 0 & 0 & 0 & -1 & \frac{3}{2} & 0\\
0 & \frac{3}{4} & 0 & -1 & 0 & 0 & 0 & \frac{1}{2}
  \end{pmatrix},
  \label{choimatrix300}
\end{eqnarray}
The Choi matrix $C_{\Phi_{\frac{3}{4},-2}}$ has at least one negative eigenvalues and thus it does not represent a positive semidefinite matrix. Hence $\Phi_{\frac{3}{4},-2}$ is a positive but not completely positive map.\\
Next our task is to show that $C_{\Phi_{\frac{3}{4},-2}}$ act as witness operator and for this it is sufficient to show that there exist at least one entangled states described by the density operator $\rho_{e}$ for which $Tr(C_{\Phi_{\frac{3}{4},-2}} \rho_{e})<0$. Then we can say that the entangled state will be detected by $C_{\Phi_{\frac{3}{4},-2}}$.\\
Let us consider a quantum state described by the density operator $\rho_{b}$ which is given by
\begin{eqnarray}
\rho_{b}= \frac{1}{1+7b}
 \begin{pmatrix}
  b & 0 & 0 & 0 & 0 & b & 0 & 0 \\
  0 & b & 0 & 0 & 0 & 0 & b & 0 \\
  0 & 0 & b & 0 & 0 & 0 & 0 & b\\
0 & 0 & 0 & b & 0 & 0 & 0 & 0 \\
0 & 0 & 0 & 0 & \frac{1+b}{2} & 0 & 0 & \frac{\sqrt{1-b^{2}}}{2}\\
b & 0 & 0 & 0 & 0 & b & 0 & 0\\
0 & b & 0 & 0 & 0 & 0 & b & 0\\
0 & 0 & b & 0 & \frac{\sqrt{1-b^{2}}}{2} & 0 & 0 & \frac{1+b}{2}
  \end{pmatrix}
  \label{pptes}
\end{eqnarray}
where the state parameter satisfies $0\leq b\leq 1$. The state $\rho_{b}$ is shown to be a bound entangled state by range criterion \cite{phorodecki}.\\
We are now in a position to show the utility of the operator $C_{\Phi_{\frac{3}{4},-2}}$ in the detection of entanglement. To accomplish this task, we calculate $Tr(C_{\Phi_{\frac{3}{4},-2}} \rho_{b})$, which is given by
\begin{eqnarray}
Tr(C_{\Phi_{\frac{3}{4},-2}} \rho_{b})=\frac{b-1}{4(1+7b)}<0
\label{witnesscond1}
\end{eqnarray}
Thus the bound entangled state $\rho_{b}$ detected by the witness operator $C_{\Phi_{\frac{3}{4},-2}}$.
\subsection{Detection of Negative Partial Transpose Entangled State}
Let us consider a quantum state described by the density operator $\rho_{NPT}$ which is given by
\begin{eqnarray}
\rho_{NPT}= \frac{1}{3}
 \begin{pmatrix}
  1 & 0 & 0 & 0 & 0 & 1 & 0 & 0 \\
  0 & 0 & 0 & 0 & 0 & 0 & 0 & 0 \\
  0 & 0 & 0 & 0 & 0 & 0 & 0 & 0\\
0 & 0 & 0 & 0 & 0 & 0 & 0 & 0 \\
0 & 0 & 0 & 0 & 0 & 0 & 0 & 0\\
1 & 0 & 0 & 0 & 0 & 1 & 0 & 0\\
0 & 0 & 0 & 0 & 0 & 0 & 0 & 0\\
0 & 0 & 0 & 0 & 0 & 0 & 0 & 1
\end{pmatrix}
\label{nptes}
\end{eqnarray}
It can be easily shown that the state (\ref{nptes}) represent a negative partial transpose entangled state. Now, our task is to construct a witness operator which can detect it. To accomplish this task, let us start with the positive input matrix which is given by
\begin{eqnarray}
A_{2}= \begin{pmatrix}
 3 & \frac{1}{3} \\
 \frac{1}{9} & 2
\end{pmatrix}
\end{eqnarray}
Considering $\beta=-\gamma (\gamma>0)$ and applying the map on $A_{2}$, we get the output matrix in the form
\begin{eqnarray}
\Phi_{\alpha,-\gamma}(A_{2})=
  \begin{pmatrix}
  6\alpha - \frac{\gamma}{2}& 0 & \frac{4\alpha}{9} & 0  \\
     0  &  6\alpha & -\frac{\gamma}{2} & \frac{4\alpha}{9}   \\
      \frac{4\alpha}{9} & -\frac{\gamma}{2} & 4\alpha  & 0 \\
    0 & \frac{4\alpha}{9} & 0 & 4\alpha - \frac{\gamma}{2}
  \end{pmatrix}
  \label{mapmat400}
\end{eqnarray}
The map $\Phi_{\alpha,-\gamma}$ will be positive map if $\gamma>0$ and $\alpha\geq \frac{9\gamma}{90-2\sqrt{27}}$.
In the next step, we fix $\gamma=1$ and then choose a value of $\alpha$ from the interval $\alpha\geq \frac{9}{90-2\sqrt{27}}$. Taking $\alpha=\frac{1}{8}$, the matrix given in (\ref{mapmat400}) reduces to
\begin{eqnarray}
\Phi_{\frac{1}{8},-1}(A_{2})=
  \begin{pmatrix}
  \frac{1}{4}& 0 & \frac{1}{18} & 0  \\
     0  &  \frac{3}{4} & \frac{-1}{2} & \frac{1}{18}   \\
      \frac{1}{18} & \frac{-1}{2} & \frac{1}{2} & 0 \\
    0 & \frac{1}{18} & 0 & 0
  \end{pmatrix}
  \label{mapmat301}
\end{eqnarray}
Therefore, the particular form of the map $\Phi_{\frac{1}{8},-1}$ represent a positive map.  Using this positive map, we can construct the Choi matrix as
\begin{eqnarray}
C_{\Phi_{\frac{1}{8},-1}}=
 \begin{pmatrix}
  \frac{-1}{4} & 0 & 0 & 0 & \frac{-1}{2} & 0 & \frac{1}{8} & 0 \\
  0 & \frac{1}{4} & \frac{-1}{2} & 0 & 0 & 0 & \frac{-1}{2} & \frac{1}{8} \\
  0 & \frac{-1}{2} & 0 & 0 & \frac{1}{8} & \frac{-1}{2} & 0 & 0\\
0 & 0 & 0 & \frac{-1}{2} & 0 & \frac{1}{8} & 0 & \frac{-1}{2} \\
\frac{-1}{2} & 0 & \frac{1}{8} & 0 & \frac{-1}{2} & 0 & 0 & 0\\
0 & 0 & \frac{-1}{2} & \frac{1}{8} & 0 & 0 & \frac{-1}{2} & 0\\
\frac{1}{8} & \frac{-1}{2} & 0 & 0 & 0 & \frac{-1}{2} & \frac{1}{4} & 0\\
0 & \frac{1}{8} & 0 & \frac{-1}{2} & 0 & 0 & 0 & \frac{-1}{4}
  \end{pmatrix},
  \label{choimatrix500}
\end{eqnarray}
The Choi matrix $C_{\Phi_{\frac{1}{8},-1}}$ has at least one negative eigenvalues and thus it does not represent a positive semidefinite matrix. Hence $\Phi_{\frac{1}{8},-1}$ is a positive but not completely positive map.\\
We will now show that $C_{\Phi_{\frac{3}{4},-2}}$ act as witness operator and it detect the state (\ref{nptes}).
To detect the state described by the density operator $\rho_{NPT}$, we calculate $Tr(C_{\Phi_{\frac{3}{4},-2}} \rho_{NPT})$, which is given by
\begin{eqnarray}
Tr(C_{\Phi_{\frac{1}{8},-1}} \rho_{NPT})=\frac{-1}{6}<0
\label{witnesscond1}
\end{eqnarray}
Thus the negative partial transpose entangled state $\rho_{NPT}$ detected by the witness operator $C_{\Phi_{\frac{1}{8},-1}}$.
\section{Conclusion}
To summarize, we have constructed a map which is applied on $n\times n$ matrix and as a result, we obtain $n^{2}\times n^{2}$ matrix at the output. The mapping constructed here is general and work for higher order matrices also. But to simplify the discussion, we have taken $n=2$ and then showed that the map is positive under certain conditions. Further, we have shown that the constructed map can never be completely positive and also obtained the conditions for which the map is positive but not completely positive. Lastly, we have discussed that the Choi matrix constructed from the positive map can act as a witness operator and take part in the detection of bound entangled state and negative partial transpose entangled state.

\end{document}